# Long-Range Hydrodynamic Response of Particulate Liquids and Liquid-Laden Solids[*]

Haim Diamant[†]

School of Chemistry, Beverly & Raymond Sackler Faculty of Exact Sciences,
Tel Aviv University, Tel Aviv 69978, Israel

**Abstract**

In viscous particulate liquids, such as suspensions and polymer solutions, the large-distance steady-state flow due to a local disturbance is commonly described in terms of hydrodynamic screening — beyond a correlation length $\xi$ the response drops from that of the pure solvent, characterized by its viscosity $\eta_0$, to that of the macroscopic liquid with viscosity $\eta > \eta_0$. For cases where $\eta \gg \eta_0$ we show, based on general conservation arguments, that this screening picture, while being asymptotically correct, should be refined in an essential way. The crossover between the microscopic and macroscopic behaviors occurs gradually over a wide range of distances, $\xi < r < (\eta/\eta_0)^{1/2}\xi$. In liquid-laden solids, such as colloidal glasses, gels and liquid-filled porous media, where $\eta \to \infty$, this intermediate behavior takes over the entire large-distance response. The intermediate flow field, arising from the effect of mass displacement rather than momentum diffusion, has several unique characteristics: (i) It has a dipolar shape with a $1/r^3$ spatial decay, negative transverse components, and vanishing angular average. (ii) Its amplitude depends on the liquid properties through $\eta_0$ and $\xi$ alone; thus, in cases where $\xi$ is fixed by geometry (e.g., for particulate liquids tightly confined in solid matrices), the large-distance response is independent of particle concentration. (iii) The intermediate field builds up non-diffusively, with a distance-independent relaxation rate, making it dominant at large distances before steady state has been reached. We demonstrate these general properties in three model systems.

---

**Introduction**

Many liquids encountered in daily life, industrial products and biology are not pure molecular liquids but contain mesoscopic particles, macromolecules, or amphiphilic structures.[1] Prime examples are colloidal suspensions[2] and polymer solutions.[3] The embedded structures increase the macroscopic viscosity of the liquid, as was first realized by Einstein for a dilute colloidal suspension.[4] The viscosity may increase by many orders of magnitude and even diverge when the macroscopic system becomes a solid, as in colloidal glasses and polymer gels.

Such particulate liquids may, in general, be non-Newtonian, viscoelastic materials. In the current work, however, we consider sufficiently slow perturbations, the response to which can be assumed purely viscous. In this limit, and at steady state, a point force $\mathbf{f}$ applied to an incompressible particle-free liquid at the origin creates a flow velocity $\mathbf{v}^{(0)}(\mathbf{r})$ at position $\mathbf{r}$ according to[5]

$$v_i^{(0)}(\mathbf{r}) = G_{ij}(\mathbf{r}) f_j, \qquad G_{ij}(\mathbf{r}) = \frac{1}{8\pi\eta_0 r}\left(\delta_{ij} + \frac{r_i r_j}{r^2}\right), \qquad (1)$$

where $\mathbf{G}$ is the Oseen tensor, $\eta_0$ being the viscosity of the pure, particle-free liquid. (Here and throughout this article we sum over repeated indices.) The slow spatial decay of the hydrodynamic response in Eq. (1) implies that the dynamics in the presence of embedded structures involve long-range and many-body correlations. Nevertheless, if the macroscopic system is still a viscous liquid, these complicated effects are expected to amount at sufficiently large distances to an effective response similar to Eq. (1),

$$v_i(\mathbf{r}) = G_{ij}^{\text{eff}}(\mathbf{r}) f_j, \qquad G_{ij}^{\text{eff}}(\mathbf{r}) = \frac{1}{8\pi\eta r}\left(\delta_{ij} + \frac{r_i r_j}{r^2}\right), \qquad (2)$$

with some macroscopic viscosity $\eta$. (If this had not been true, one would not have been allowed to apply the Stokes equation to macroscopic flows of particulate liquids.) The flow field of Eqs. (1) and (2) is depicted in Fig. 1A. In cases where the macroscopic system is a solid (e.g., colloidal glasses, polymer gels, biological actin networks) $\eta \to \infty$ and the effective response in Eq. (2) evidently vanishes.

The effect embodied in Eq. (2) is sometimes referred to as *hydrodynamic screening*.[3,6] It emerges in various effective-medium theories for the dynamics of particulate liquids (e.g., Refs. 3,7-11). The crossover between the microscopic response, Eq. (1), and the macroscopic one, Eq. (2), occurs beyond a certain correlation length $\xi$, which depends on the particular properties of the dissolved structures. (For example, in polymer solutions it is comparable to the thermodynamic correlation length of the polymer assembly,[3] and in electrolytes $\xi$ is the Debye screening length.[10]) This screening picture is schematically illustrated in Fig. 2. The current work is focused on the crossover region, represented in Fig. 2 by a dashed line. We will show that the hydrodynamic response in this region is qualitatively different from those in the microscopic and macroscopic ones. Moreover, as $\eta$ increases, not only is the macroscopic response suppressed in magnitude, but it is also pushed to larger distances while the intermediate region becomes increasingly broad.

It should be stressed that the emergence of a long-range flow field, qualitatively different from Eqs. (1) and (2), is not a new observation. The Brinkman model for flow through porous media[11] yields a far flow of dipolar shape, decaying as $1/r^3$, as was noted by Durlofsky and Brady.[12] This was further emphasized by Long and Ajdari[10] in a study of hydrodynamic interactions in gels and driven electrolytes based on Brinkman-like equations. Using a mean-field calculation and numerical study of lattice-suspensions, Mucha et al.[13] found that the hydrodynamic interaction in concentrated suspensions decayed as $1/r^3$. Finally, we mention a somewhat overlooked point concerning dynamical theories of semidilute polymer solutions.[3] In Fourier space and in the limit of small wavevectors these theories yield an effective response similar to that of a Brinkman liquid. Thus, upon inversion back to real space, they too exhibit a far dipolar flow. (The origin of this dipolar field will be clarified below.)

The aim of the current work is to generalize these results, hopefully putting them on more fundamental grounds. We thereby draw several new conclusions regarding the unusual characteristics of the hydrodynamic response in the intermediate regime.

## Momentum Conservation and the Asymptotic Monopolar Flow

The equations of hydrodynamics derive from conservation laws for slow coarse-grained variables. In Newtonian liquids these are the density fields of momentum, mass, and energy. All the systems considered here are assumed to have sufficiently fast heat diffusion, such that the liquid is isothermal and does not conserve energy.

Applying a point force to a liquid element introduces a source of momentum (momentum monopole) at that element. The Navier-Stokes equation describes the propagation of momentum from the source throughout the liquid. In the limit of an incompressible liquid longitudinal modes are suppressed and only the transverse components diffuse away, with a diffusion coefficient equal to the kinematic shear viscosity of the liquid, $v_0 = \eta_0/\rho_0$, $\rho_0$ being the liquid mass density. This diffusive signal is the physical origin of the Oseen tensor, Eq. (1). Its $1/r$ spatial decay arises from the steady-state distribution of a three-dimensional random walk (e.g., the fundamental solution of the Laplace equation), while its angular dependence is dictated by the incompressibility condition, i.e., the requirement $\partial_i G_{ij} = 0$. Consequently, the flow shown in Fig. 1A is a monopolar field, decaying slowly as $1/r$ and having all flow lines roughly in the same direction as that of the vector source.

From another point of view, the total momentum flux through any closed surface containing the source must be constant. The local momentum flux is just the liquid stress tensor. Thus, the stress must decay with distance as $1/r^2$ and, since flow velocity is related to stress by one spatial derivative, we have $|\mathbf{G}| \sim 1/r$. It is therefore clear that, regardless of the complicated structures that the system may contain, as long as it is a momentum-conserving incompressible liquid, its large-distance steady-state response must be described by the effective Oseen tensor of Eq. (2). In other words, the screening description is asymptotically exact.

The physical picture is clear too.[1] Over distances from the source much smaller than $\xi$ the probability to encounter a particle is small, and momentum diffuses through the pure liquid, as described by Eq. (1) with the prefactor $\eta_0^{-1}$. At distances $r \mathrel{>\!\!\sim} \xi$ momentum is imparted to the much less mobile particles and diffusion is impeded. At yet larger distances particle motion becomes appreciable and momentum is exchanged between the particles and the host liquid. The diffusion through the mixed system, involving the conservation of total momentum, is then described by Eq. (2) with the smaller prefactor $\eta^{-1} < \eta_0^{-1}$.

None of the arguments given above implies, however, that at distances $r > \xi$ momentum diffusion is necessarily the dominant effect in the flow response. This will be clarified in the next section.

If the liquid pervades a solid matrix, momentum will be imparted to the solid and, subsequently, no matter how tenuous the solid structure may be, transferred quickly (with the sound velocity of the solid) to the container. Thus, such liquid-laden solids (colloidal glasses, gels, liquid-filled porous media) do not conserve momentum. Momentum will diffuse from the source to a nearby solid obstacle, along a typical distance $\xi$, and be lost. At large distances contributions from the momentum monopole are therefore exponentially small in $r/\xi$.

**Mass Conservation and the Intermediate Dipolar Flow**

To see that the screening mechanism described above is not the whole story, consider the following simple thought experiment. Let an incompressible liquid pervade a porous medium and place a source of liquid mass at the origin. Obviously, regardless of the structure of the solid matrix or loss of liquid momentum to it, the source will create a long-range flow velocity, $\mathbf{v}_1 \sim (1/r^2)\hat{\mathbf{r}}$, to conserve liquid mass. Now add a sink of liquid at position $-\mathbf{l}$, thereby creating a mass dipole pointing in the $+\mathbf{l}$ direction. The resulting large-distance liquid velocity will be $v_i \sim -l_j \partial_j v_{1i} \sim -r^{-3}(\delta_{ij} - 3r_i r_j/r^2)l_j$. The same will hold, of course, in any liquid or liquid-laden solid.

Any finite-size disturbance displaces liquid mass and thus introduces a mass dipole pointing in the direction of the displacement. For a particle of radius $a$, moving through an otherwise empty liquid, $l \sim a$ and the velocity inferred above is the well known dipolar term of the flow due to single-particle motion.[5] Yet, in the case of many particles or a matrix, with correlation length $\xi$, we can localize the applied force in an arbitrarily small volume and yet have liquid mass displaced over $l \sim \xi$. (In the language of polymer dynamics we have a correlation "blob" of size $\xi$ that acts as an effective particle, regardless of the size of the original disturbance.) Dimensionally, the strength of the resulting mass dipole per unit force must be proportional to $\xi^2/\eta_0$. We thus have, in summary,

$$\mathbf{G}^{\text{eff}} = \mathbf{G}^{\text{m}} + \mathbf{G}^{\text{d}},$$

$$G_{ij}^{\text{m}}(\mathbf{r}) = \frac{1}{8\pi\eta r}\left(\delta_{ij} + \frac{r_i r_j}{r^2}\right), \quad G_{ij}^{\text{d}}(\mathbf{r}) \sim -\frac{\xi^2}{\eta_0 r^3}\left(\delta_{ij} - 3\frac{r_i r_j}{r^2}\right). \tag{3}$$

While both the monopolar and dipolar terms in Eq. (3) are valid for $r \gg \xi$, each is dominant in a different region — $\mathbf{G}^{\text{d}}$ dominates in the intermediate range $\xi \ll r \ll \xi(\eta/\eta_0)^{1/2}$, and $\mathbf{G}^{\text{m}}$ in the asymptotic large-distance region, $r \gg \xi(\eta/\eta_0)^{1/2}$ (as demanded by momentum conservation). The intermediate range becomes increasingly wide as the effective viscosity of the particulate liquid gets larger. The three regimes are schematically summarized in Fig. 2. They are strictly applicable only at steady state; before steady state has been reached throughout the system, the picture is qualitatively different, as will be discussed below.

The intermediate flow field is depicted in Fig. 1B. It has several noteworthy characteristics. First, we note a few mathematical properties. The flow field decays with distance as $1/r^3$, faster than the usual monopolar field but still only algebraically. Its dipolar shape involves circulation flows which make transverse components negative, e.g., $G_{xx}^d(r\hat{\mathbf{y}}) \sim -\xi^2/(\eta_0 r^3) < 0$. Thus, at positions transverse to the direction of the disturbance the flow points in the opposite direction. (See Fig. 1B.) This implies that a pair of particles positioned in this way will experience "anti-drag". The field is not only divergence-free, $\partial_i G_{ij}^d = 0$, as demanded by incompressibility, but is also trace-free, $G_{ii}^d = 0$, and Laplacian-free, $\partial_{kk} G_{ij}^d = 0$. Finally, the angular average of each component vanishes, $\int d\Omega G_{ij}^d = 0$. An important implication of the last property is that any calculation involving a pre-averaging approximation for the Oseen tensor (as commonly assumed, e.g., in studies of polymer dynamics[3,14]) will not be affected by the dipolar contribution.

Second, we notice that the dipolar field depends on the properties of the liquid only through $\xi$ and the pure-solvent viscosity $\eta_0$. Hence, in systems where $\xi$ is dictated by some fixed parameters we expect $\mathbf{G}^d$ to be insensitive to the properties of the particulate liquid. For example, when a suspension or a polymer solution is tightly confined within a porous matrix, the hydrodynamic correlation length will be determined by the pore width, $\xi \sim w$. Furthermore, as discussed above, the large-distance response of such liquid-laden solids is given solely by the dipolar field, $\mathbf{G}^{\text{eff}} = \mathbf{G}^d$. We thus reach the surprising conclusion that increasing the concentration of particles or polymers in such confined liquids will have no effect on the large-distance response. This will be true, obviously, only so long as the correlation volume $\xi^3 \sim w^3$ does not contain many particles; otherwise, the regular screening of Eq. (2) will set in at a length scale smaller than $w$, leading to a concentration-dependent effective viscosity as in an unbounded liquid. The conclusion regarding the concentration independence, therefore, is limited to particle volume fractions not much larger than $a^3/w^3$.

Third, we recall that the physical origin of the dipolar field (mass displacement) is essentially different from that of the monopolar one (transverse momentum diffusion). We should expect, therefore, a qualitatively different temporal evolution for the two contributions. Specifically, if a local disturbance is applied at time $t = 0$ to a momentum-conserving liquid, the monopolar flow field will build up over large distances (small wavevector $q$) via a diffusive mode with a $q$-dependent rate, $\Gamma_m = \nu q^2$, $\nu = \eta/\rho$ being the effective kinematic viscosity of the liquid. By contrast, the mass-displacement perturbation propagates over large distances with velocity $c_s$, the sound velocity in the liquid, which in the limit of incompressibility is assumed infinite. Hence, the time it should take this field to build up is merely the time needed for transverse momentum to diffuse a distance $\xi$ away from the source and establish the effective mass dipole. The buildup rate of the dipolar field is thus $\Gamma_d \sim \nu_0/\xi^2$, independent of $q$. Consequently, at times $\xi^2/\nu_0 < t < L^2/\nu$ ($L$ being the system size), the monopolar field has not yet reached steady state while the dipolar one already has. Before steady state is attained throughout the system, therefore, it is the dipolar contribution that actually dominates the flow response at asymptotically large distances, $r \gg \xi(\nu/\nu_0)^{1/2}$, even in momentum-conserving liquids. In addition, we conclude that in liquid-laden solids the temporal hydrodynamic response builds up quickly throughout the system with a distance-independent rate (so long as the system size does not exceed $c_s \xi^2/\nu_0$).

It is instructive to see how the general properties inferred above are manifest in concrete solvable examples. This is the purpose of the next three sections.

**Example 1: Dilute Suspension of Hard Spheres**

Consider a colloidal suspension containing a volume fraction $\phi$ of hard spheres of radius $a$, suspended in a host liquid of viscosity $\eta_0$. In the dilute limit, $\phi \ll 1$, we expect that the hydrodynamic response at large distances will be that of Eq. (2), with the effective viscosity of a dilute suspension as calculated by Einstein,[4]

$$\eta = \eta_0 \left(1 + \frac{5}{2}\phi\right). \tag{4}$$

At smaller distances, yet still much larger than $a$, we expect to find the dipolar term inferred in the preceding section, with $\xi$ proportional to $a$ (the only length in the problem) and increasing with $\phi$. We will now prove that these statements are indeed correct. The calculation is straightforward and was probably done before. However, we have not been able to find it in the literature and therefore give it here in some detail. (The same result is obtained as the dilute limit of the much more complicated Beenakker-Mazur theory of colloidal suspensions.[15])

We begin with a particle-free liquid. A point force $\mathbf{f}$ is applied to the liquid at the origin. It creates a flow field whose velocity at point $\mathbf{r}$ is given by

$$v_i^{(0)}(\mathbf{r}) = G_{ij}(\mathbf{r}) f_j, \tag{5}$$

where $\mathbf{G}$ is the Oseen tensor of Eq. (1). Let us now add at position $\mathbf{r}'$ a particle, which obstructs the flow $\mathbf{v}^{(0)}$. Since the particle is force- and torque-free, the obstruction is accounted for, to a leading moment, by a force dipole (stresslet), $\mathbf{S}(\mathbf{r}')$. The force dipole should be proportional to the variation of the original flow velocity over the volume now occupied by the particle, i.e., to spatial derivatives of $\mathbf{v}^{(0)}$ at the position $\mathbf{r}'$. The exact relation is provided by Faxen's second law,[2]

$$S_{ij} = \frac{10\pi}{3} \eta_0 a^3 \left(1 + \frac{a^2}{10}\partial_{kk}\right)\left(\partial_i v_j^{(0)} + \partial_j v_i^{(0)}\right). \tag{6}$$

This force dipole, located at $\mathbf{r}'$, creates an extra flow velocity at $\mathbf{r}$ according to

$$\delta v_i(\mathbf{r}, \mathbf{r}') = S_{jk}(\mathbf{r}')\partial_j G_{ki}(\mathbf{r} - \mathbf{r}'). \tag{7}$$

Equation (7) gives the correction to the particle-free flow velocity $\mathbf{v}^{(0)}$ due to the presence of a single particle at $\mathbf{r}'$. In the dilute limit under consideration higher-order (two-particle) terms are negligible. Thus, to sum up the contributions from all particles at all possible positions we can use the uniform probability density of finding a particle centered at $\mathbf{r}'$, which is equal to $\phi/(4\pi a^3/3)$. This leads to the average correction,

$$\langle \delta v_i(\mathbf{r}) \rangle = \frac{3}{4\pi a^3} \phi \int d^3 r' \, S_{jk}(\mathbf{r}')\partial_j G_{ki}(\mathbf{r} - \mathbf{r}'). \tag{8}$$

The convolution in Eq. (8) is easily performed in Fourier space, where the Oseen tensor reads

$$\tilde{G}_{ij}(\mathbf{q}) = \frac{1}{\eta_0 q^2}\left(\delta_{ij} - \frac{q_i q_j}{q^2}\right). \tag{9}$$

Substitution in Eqs. (5), (6) and (8) leads to

$$\tilde{v}_i(\mathbf{q}) = \tilde{v}_i^{(0)} + \delta\tilde{v}_i = \tilde{G}_{ij}^{\text{eff}}(\mathbf{q})f_j, \quad \tilde{G}_{ij}^{\text{eff}}(\mathbf{q}) = \left[\left(1-\frac{5}{2}\phi\right) + \alpha\phi a^2 q^2\right]\tilde{G}_{ij}(\mathbf{q}), \tag{10}$$

where $\alpha$ is a numerical coefficient to be discussed below. Upon inversion back to real space, we get

$$G_{ij}^{\text{eff}}(\mathbf{r}) = \frac{1}{8\pi\eta r}\left(\delta_{ij} + \frac{r_i r_j}{r^2}\right) - \frac{\xi^2}{4\pi\eta_0 r^3}\left(\delta_{ij} - 3\frac{r_i r_j}{r^2}\right). \tag{11}$$

Equation (11) has the same form as the response inferred earlier on general grounds, Eq. (3), with $\eta^{-1} = \eta_0^{-1}(1-5\phi/2)$, in accord (to the assumed linear order in $\phi$) with Einstein's effective viscosity [Eq. (4)], and $\xi = (\alpha\phi)^{1/2}a$.

The analysis described above yields $\alpha = 1/4$. This coefficient is incorrect, in fact, since another term of order $q^2\tilde{\mathbf{G}}$ in Eq. (10), arising from a third force moment, has been disregarded. Direct calculation of the correct coefficient lies beyond the scope of the current work, which aims to demonstrate the general form of the large-distance response. Nonetheless, we can find $\alpha$ by taking the dilute limit of the Beenakker-Mazur theory.[15,16] To leading order in $\phi$ and $qa$ this theory yields a result identical to Eq. (11) with $\alpha = 5/6$.

**Example 2: Suspension Confined between Two Plates**

Arguably the simplest example of a liquid-laden solid is provided by two parallel plates, a distance $w$ apart, confining a slab of liquid of viscosity $\eta_0$. Since such a slab does not conserve momentum, as discussed above, its large-distance hydrodynamic response should be governed by a mass-displacement dipolar field. In the current example large distances are attainable only along the two axes parallel to the confining surfaces. Hence, the far flow field due to a disturbance directed parallel to the plates is expected to be that of a two-dimensional mass dipole,

$$G_{ij}^{\text{eff}}(\mathbf{r}) \sim -\frac{\xi}{\eta_0 r^2}\left(\delta_{ij} - 2\frac{r_i r_j}{r^2}\right), \tag{12}$$

which is the two-dimensional analogue of $\mathbf{G}^d$ of Eq. (3). The flow due to a point force in a particle-free liquid confined between two plates was calculated in Ref. 17. For a point force located at the midplane and directed parallel to the plates the exact solution for the far flow field is identical to Eq. (12) with a prefactor of $3/(32\pi)$ and $\xi = w$, the slab width.

The interesting point concerning this system is that $\xi$ is fixed by the confined geometry. Thus, if the slab contains suspended particles (so long as their volume fraction is not too high, as discussed earlier), the large-distance hydrodynamic response will still be given by Eq. (12) with $\xi = w$, regardless of particle concentration. Consider, for example, a layer of particles suspended at the midplane between the plates.[18-20] Hydrodynamic analysis of such a quasi-two-dimensional suspension,[19] along lines similar to those of the preceding section, indeed yields a large-distance response independent of particle volume fraction, to linear order in $\phi$. These conclusions —the dipolar shape of the far flow [Eq. (12)] with its $1/r^2$ decay and negative transverse component ("anti-drag"), its independence of particle concentration— have all been confirmed to high precision in recent experiments.[18-20]

**Example 3: Time-Dependent Response of a Liquid-Laden Porous Medium**

Finally, to demonstrate the non-diffusive temporal buildup of the dipolar flow field, we consider the hydrodynamic response of a liquid embedded in a porous solid matrix using the Brinkman theory.[11,12] This effective-medium model accounts for the effect of the solid matrix by adding to the Navier-Stokes equation for the liquid a phenomenological momentum-loss term. The resulting equations for an incompressible liquid are

$$\rho_0(\partial_t v_i + v_j \partial_j v_i) = -\partial_i p + \eta_0 \partial_{jj} v_i - \rho_0 \Gamma v_i + f_i,$$
$$\partial_i v_i = 0,$$
(13)

where $\mathbf{v}(\mathbf{r},t)$ is the liquid velocity at position $\mathbf{r}$ and time $t$, $p(\mathbf{r},t)$ the pressure, $\mathbf{f}(\mathbf{r},t)$ the force density applied to the liquid, $\rho_0$ the liquid mass density, $\eta_0$ its viscosity, and $\Gamma$ the rate of momentum loss to the solid matrix. When the inertial terms in Eq. (13) are neglected, the Brinkman theory provides an interpolation between the Stokes equation, which must hold at sufficiently short distances (shorter than $\xi = (\nu_0/\Gamma)^{1/2}$, $\nu_0 = \eta_0/\rho_0$ being the kinematic viscosity), and Darcy's law, which describes macroscopic liquid permeation through a porous medium with permeability $\xi^2 = \nu_0/\Gamma$.

We consider an initially stationary liquid, $\mathbf{v}(\mathbf{r},0) = 0$, to which a step-function point force is applied at $t = 0$,

$$\mathbf{f}(\mathbf{r},t) = \mathbf{F}\delta(\mathbf{r})\Theta(t),$$
(14)

and ask what the time-dependent flow velocity will be at large distances. Assuming a small perturbation we may neglect the nonlinear term in Eq. (13). It is then straightforward to solve the problem defined by Eqs. (13) and (14) using a spatial Fourier transform and temporal Laplace transform. This yields, for $r \gg \xi$,

$$v_i(\mathbf{r},t) = G_{ij}^{\mathrm{eff}}(\mathbf{r},t) F_j,$$
$$G_{ij}^{\mathrm{eff}}(\mathbf{r},t) \simeq -\frac{\xi^2}{4\pi\eta_0 r^3}\left(\delta_{ij} - 3\frac{r_i r_j}{r^2}\right)(1 - e^{-\Gamma t}).$$
(15)

(The additional monopolar field is exponentially small in $r/\xi$.) Thus, the dipolar flow field builds up non-diffusively, with a distance-independent rate. Since $\Gamma = \nu_0/\xi^2$, Eq. (15) conforms precisely with the general temporal behavior inferred earlier.

**Discussion**

We have suggested in this article a new perspective at the large-distance hydrodynamic response of structured liquids and liquid-laden solids, based on general conservation arguments rather than a detailed specific theory. This perspective provides an insight into the mechanisms of hydrodynamic screening, highlighting the emergence of a long-range intermediate flow field of unique spatial and temporal properties. Several of these properties have been previously established in more specific models, e.g., the $1/r^3$ spatial decay[10,12,13] (or $1/r^2$ in quasi-two-dimensional suspensions[17-19]), the "anti-drag",[18,19] and the cancellation upon pre-avaraging.[10,14] These properties have been shown here to arise from more general, model-independent principles. In addition, we have demonstrated a few properties which were not appreciated before, in particular, the non-diffusive dynamics of the intermediate flow field and its independence of particle concentration within solid matrices. We have focused on the flow field due to an applied point force, yet the results evidently bear upon such additional issues as the effective large-distance hydrodynamic interaction between particles, the effective pair-mobility coefficients, and (through the Einstein relation) the effective pair-diffusion coefficients.

Another consequence of the current analysis is a set of strict limits against which the asymptotic validity of various detailed theories or simulations may be examined. For example, based on the arguments given above, we can infer that the phenomenological Brinkman theory for flow in porous media must be asymptotically exact at large distances. This is because (a) this theory [i.e., Eq. (13)] produces the correct mass-dipolar flow field at large distances, and (b) the coefficient of the extra momentum-loss term introduced in the theory is unambiguously determined by the macroscopic permeability of the medium to properly reproduce Darcy's law. Reference 12 presents a detailed numerical analysis of the validity of the Brinkman theory, in which one sees that the numerical results and phenomenological theory indeed converge at sufficiently large distances for all volume fractions of the solid. Opposite examples are provided by theories and analyses of experimental data, pertaining to momentum-conserving structured liquids, which yield far flow fields that decay faster than $1/r$.[13,21,22] While such results may be valid over a certain intermediate range, they have to be asymptotically incorrect.[23]

The effects predicted here regarding the intermediate dipolar field are not weak and, hence, should be experimentally observable. The traditional dynamic scattering techniques are problematic in this respect, since they involve thermodynamic and spatial averages which, as discussed above, may annul the effect of the dipolar flow field. Recently, however, the focus has been shifting to real-space, more direct techniques involving manipulation and tracking of individual particles. The application of such techniques to quasi-two-dimensional suspensions[18-20] has already confirmed the conclusions drawn in Example 2 above. Our main claim here is that the peculiar hydrodynamic effects observed in those studies, in fact, have implications for a much wider variety of systems. They could be observed, for example, in the low-frequency correlated motion of particles embedded in a semi-dilute polymer solution or a porous matrix. Our discussion has been restricted to purely viscous (zero-frequency) effects. An important question that naturally arises is how these ideas bear upon the frequency-dependent viscoelastic response of particulate liquids.

**Acknowledgments**

The ideas presented in this article have emerged from numerous discussions with colleagues during the past few years. The author is particularly thankful to Armand Ajdari, David Andelman, Michael Cates, Benny Davidovitch, Rony Granek, Alexander Grosberg, Sriram Ramaswamy, David Reichman, Stuart Rice, Tsvi Tlusty, and Tom Witten. This work was supported by the Israel Science Foundation under Grant nos. 77/03 and 588/06.

**Figure Captions**

**Figure 1**: Cross-sections of large-distance flow fields due to a local disturbance. (A) Flow field due to a point force (momentum monopole) located at the center and directed to the right [Eq. (2)]. This monopolar field governs the asymptotically large-distance response of momentum-conserving liquids. (B) Flow field due to a mass dipole located at the center and pointing to the right [$\mathbf{G}^d$ of Eq. (3)]. This dipolar field is dominant at intermediate distances in momentum-conserving liquids and over all large distances in liquid-laden solids.

**Figure 2**: Schematic diagram of the three hydrodynamic-response regimes using a logarithmic scale. The microscopic and macroscopic responses (solid lines) are governed by monopolar flow fields (see Fig. 1A) decaying with distance as $1/r$. In the crossover region (dashed line) the flow field is dipolar (Fig. 1B) with a $1/r^3$ decay. The upper limit of the crossover region advances to increasingly larger distances as the effective viscosity of the liquid increases.

**Figures**

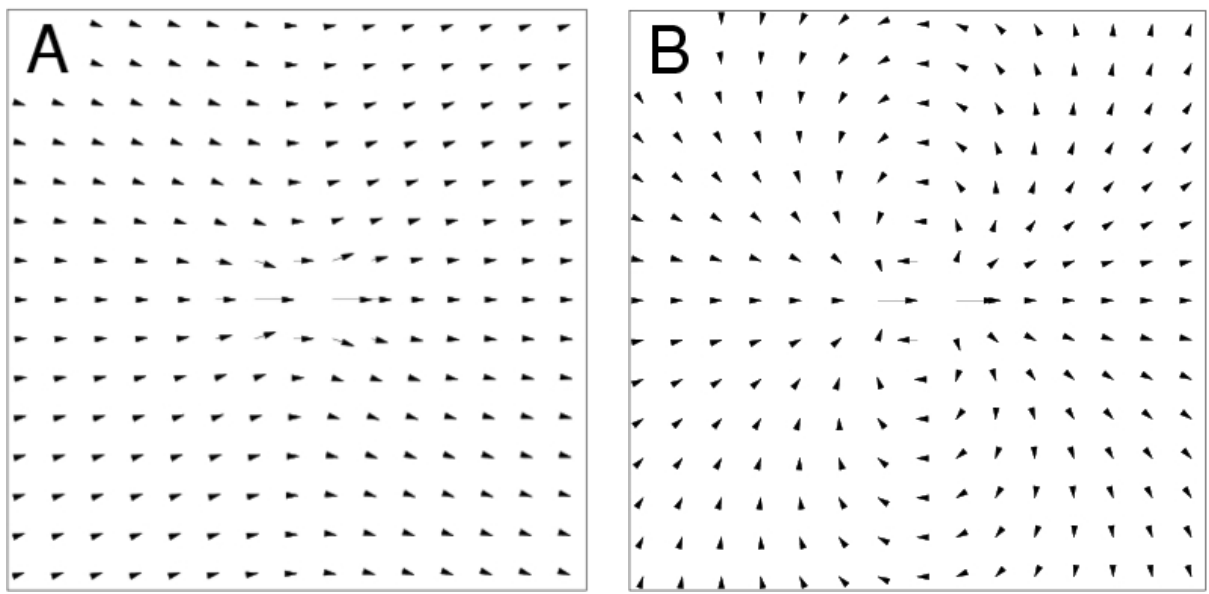

**Figure 1**

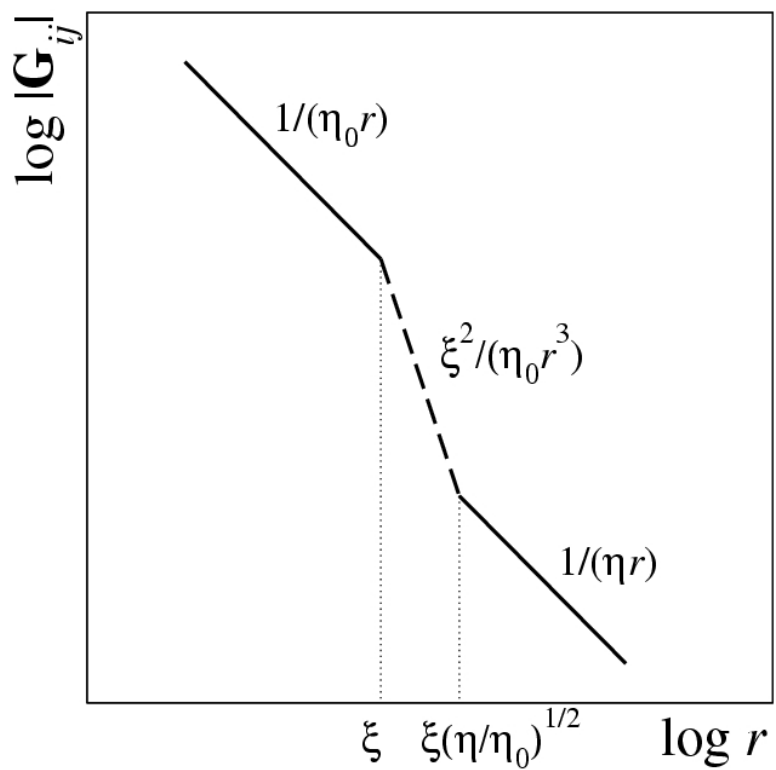

**Figure 2**